\documentclass[aps,prl,twocolumn,groupedaddress]{revtex4-2}

\usepackage{amsmath}
\usepackage{amssymb}
\usepackage{graphicx}
\usepackage[colorlinks=true, allcolors=blue]{hyperref}
\usepackage[dvipsnames]{xcolor}

\begin{document}
	
	\title{Five-point Superluminality Bounds}
	
	\author{Francesco Serra}
	\email[]{fserra2@jh.edu}
	\affiliation{Department of Physics and Astronomy, Johns Hopkins University, Baltimore, MD 21218, USA}
	
	\author{Leonardo G. Trombetta}
	\email[]{trombetta@fzu.cz}
	\affiliation{CEICO, Institute of Physics of the Czech Academy of Sciences, Na Slovance 1999/2, 182 00, Prague 8, Czechia.}
	
	\date{\today}
	
	\begin{abstract}
		We investigate how the speed of propagation of physical excitations is encoded in the coefficients of five-point interactions. This leads to a superluminality bound on scalar five-point interactions, which we present here for the first time. To substantiate our result, we also consider the case of four-point interactions for which bounds from S-matrix sum rules exist and show that these are parametrically equivalent to the bounds obtained within our analysis. Finally, we extend the discussion to a class of higher-point interactions.
	\end{abstract}
	
	\maketitle

	\section{Introduction}
	
	Fundamental interactions are universally characterized, at low enough energies, by the relevant field degrees of freedom, relevant energy scales and by the symmetries that govern the dynamics. This effective field theory (EFT) characterization allows to describe low energy processes with a given precision by specifying the coefficients of a finite number of interaction operators. Faced with the vast parameter space of yet undetermined coefficients that describe fundamental interactions in nature, an important and ambitious task is to develop a systematic understanding of how key dynamical features manifest at the level of these coefficients. A remarkable example of such manifestation is given by unitary, Lorentz-invariant effective field theories that respect microcausality, i.e. (anti)commutation of operators evaluated on space-like separated regions. These theories do not admit arbitrary coefficients for large classes of low-energy interactions, see e.g. \cite{Pham:1985,Adams:2006sv,Arkani-Hamed2021a}. More in detail, in these theories one expects the S-matrix to be an analytic function of the center of mass energy invariant, aside for particle production branch cuts and simple poles, as well as possible anomalous thresholds \cite{Bros1964}, see e.g. \cite{Mizera2024} for a pedagogic review. This property, together with crossing symmetry and boundedness of the S-matrix elements implied by unitarity, allows to constrain the coefficients of certain low-energy interactions through sum rules. For two-to-two S-matrix elements, these constraints impose either the positivity or relative boundedness of the coefficients of the low energy 4-point interactions, see e.g. \cite{Nicolis:2009qm,Bellazzini2017,Bellazzini:2017fep,Bellazzini:2020cot,Tolley2021,Caron-Huot2021,Bellazzini2022,Caron-Huot:2021rmr,Chiang:2021ziz,Caron-Huot2022,Hong:2023zgm}. Beyond two-to-two S-matrix elements little is known. The crossing and analytic properties of the S-matrix are less transparent for higher-point processes \cite{Bros1972,CaronHuot2023a,CaronHuot2023}, making it more difficult to obtain sum rules \cite{Bros1986,Chandrasekaran2018}. 
	
	While the principles underpinning the S-matrix sum rules characterize both low-energy (IR) and high-energy (UV) processes, similar constraints to those extracted from the S-matrix elements can be derived without any apparent statements regarding UV physics. This, by requiring that low energy excitations should not propagate faster than light \cite{Camanho:2014apa,CarrilloGonzalez2022,Serra2022,CarrilloGonzalez2023}. Although the relation between these two different sets of bounds is not rigorously understood, their similarity can be traced back to how microcausality both determines the analyticity of the S-matrix elements and generically implies the absence of superluminal propagation \cite{Adams:2006sv,Arkani-Hamed2021a}. (See however \cite{Greene:2022uyf} for a discussion of cases in which this second connection can be invalidated.)
	
	In light of this, one can sidestep obstacles in the sum rules and make progress in charting the parameter space of interactions by studying superluminality at low energies.
	Here we follow this idea and investigate how the absence of superluminality constrains 5-point interactions, finding for the first time a two-sided bound on quintic derivative interactions.
	
	For simplicity, we consider the case of a real, shift-symmetric scalar field with derivative self-interactions as a proxy for more phenomenologically relevant systems. Bounds on some of the 4-point interaction coefficients appearing in these effective theories have already been derived through S-matrix sum rules, see e.g. \cite{Caron-Huot2021,Bellazzini2022,Tolley2021}, as well as considering superluminality at low energies \cite{CarrilloGonzalez2022}. 
	The results from S-matrix sum rules in this context can be interpreted in terms of symmetries compatible with microcausality and Lorentz invariance. In particular, at the level of four-point interactions, the S-matrix analysis implies that the Galileon symmetry of the scalar, $\phi\to\phi+a+b_\mu x^\mu\,$ \cite{Nicolis2009b}, with $a\,,b_\mu$ constants, is not compatible with the sum rules and that the Galileon invariant 4-point interactions should be subleading with respect to the Galileon breaking ones in the EFT.
	
	In the following we study superluminal propagation on the simplest possible scalar field background. We derive bounds on 4-point interactions that are parametrically equivalent to those found in the literature. We then derive the central result of this work, a bound on 5-point interactions. This bound states that also at 5-point the Galileon symmetry implies superluminal propagation. This also leads to incompatibility with microcausality, as long as no subtleties as in \cite{Greene:2022uyf} arise. We finally generalize the bounds beyond Galileon operators and extend the analysis to a class of higher-point operators.

	\section{Extracting time-shifts from the acoustic metric}
	We start by considering the following Lagrangian:
	\begin{eqnarray}
		\mathcal{L} &=& -\frac{1}{2} \Bigl\{ (\partial \phi)^2 - \frac{1}{2} c_2 (\partial \phi)^4 + c_3 (\partial \phi)^2 \square \phi + c_4 (\partial \phi)^2 E_4 \notag \\
		&&\qquad+ c_5 (\partial \phi)^2 E_5 \Bigr\}+\dots\;,
	\end{eqnarray}
	where the coefficients $c_i$ are dimensionful, e.g. $c_2 \sim \Lambda_2^{-4}$, $c_3 \sim \Lambda_3^{-3}$, $c_4 \sim \Lambda_3^{-6}$, $c_5 \sim \Lambda_3^{-9}$ (with $\Lambda_{2,3}$ some energy scales), while $E_4=(\Box\phi)^2-(\partial^2\phi)^2$ and $E_5=(\Box\phi)^3-3\Box\phi(\partial^2\phi)^2+2(\partial^2\phi)^3$. Dots indicate other operators which we assume to be suppressed by smaller coefficients and therefore negligible. In the following, we will adopt the notation $\Pi_{\mu\nu}=\partial_\mu\partial_\nu\phi\,$. In this Lagrangian the Galileon symmetry is only broken by the $c_2$ contribution. The shift-symmetry of the field, $\phi\to\phi+a\,$, is instead unbroken, leaving the scalar massless. In this theory, we consider a perturbation $\varphi$ propagating on a background ${\phi}$, and proceed by linearizing the equation of motion with respect to $\varphi$,   
    \begin{eqnarray}\label{eomlin}
        Z^{\mu\nu} \partial_\mu \partial_\nu \varphi + D^\mu \partial_\mu \varphi = 0 \, ,
    \end{eqnarray}
    where \mbox{$D^{\nu}= - c_2 \left( 4  \partial_\mu  {\phi}\Pi^{\mu\nu} \, + 2 \Box\phi \partial^\nu  {\phi} \right)$} and $Z^{\mu\nu}$ is the acoustic metric:
    \begin{eqnarray}\label{Z}
        Z_{\mu\nu} \!\! &=& \!\! \left ( 1 - c_2 (\partial  {\phi})^2 + 2 c_3 \Box\phi + 3 c_4 E_4 
        + 4c_5 E_5 \right ) \eta_{\mu\nu} 
        \notag \\	&&
        - 2 c_2 \, \partial_\mu  {\phi} \, \partial_\nu  {\phi} 
        - 2 \left( c_3 + 3 c_4 \Box\phi + 6 c_5 E_4 \right) \Pi_{\mu\nu} \notag \\
        &&+ 6 \left( c_4 + 4 c_5 \Box\phi \right) \Pi_{\mu\alpha}{\Pi^\alpha}_\nu  - 24 c_5 \, \Pi_{\mu\alpha}\Pi^{\alpha\beta}\Pi_{\beta\nu}.\,\quad
	\end{eqnarray}
    We then use an Eikonal ansatz with \mbox{$\varphi = A(x) \exp(-iS(x)/\xi)$}, where the variation of the amplitude is assumed to be much slower than that of the phase, i.e. formally $\xi \to 0$. In this geometrical optics limit, the leading order equation dictates that the wavefronts $S = const$ propagate along the null directions of the background-dependent acoustic metric 
    \begin{eqnarray} \label{eom}
        Z^{\mu\nu}(x) \, k_\mu k_\nu = 0,
    \end{eqnarray}
    where $k_\mu \equiv \partial_\mu S(x)$. This equation fully describes the changes in propagation speed of the perturbation $\varphi$. On the other hand, the $D^\mu$ term in Eq.~\eqref{eomlin} only affects the evolution of the amplitude $A(x)$, which is subleading for $\xi\to 0$. These dissipation/enhancement effects can be neglected if propagation takes place over short enough distances, as we will discuss below, see Eq.~\eqref{Lmax}. If this is the case one can just solve $Z^{\mu\nu} k_\mu k_\nu=0$. Assuming a locally diagonizable $Z^{\mu\nu}{(x)}$, there is always a local Lorentz frame where propagation is fully described by a set of 3 (potentially different) soundspeeds,
	\begin{eqnarray} \label{diagonal-disp-rel}
		\omega^2 = c_{si}^2 (x) k_i^2\;,
	\end{eqnarray}
	where the vectors $\vec{k}_i$ can in principle be functions of $x$. In the following we will assume that propagation takes place on a fixed direction $\vec{k}_i=\vec{k}=const$ and that $c_s=1+\delta c$, with $|\delta c|\ll 1$. Both of these assumptions entail a restriction on choice of background. Superluminality will then correspond to time-like $k_\mu$ and wavefronts that propagate outside the lightcone: $\eta^{\mu\nu}k_\mu k_\nu=-\omega^2+\vec{k}^2 \simeq -2\delta c{(x)} \,\vec{k}^2$, where we are neglecting $O(\delta c\,^2)$ corrections. In the following sections we will work in a weak-field regime for the background, well in the regime of validity of perturbation theory. This ensures that {the assumption} $|\delta c|\ll1$ holds in the cases that we consider. In this setup, when a signal propagates through a region of size $L$ in the background ${\phi}$, it will accumulate a phase-shift from which one can extract the following time delay or advance \cite{Maiani:1997pd}:
	\begin{eqnarray}\label{DT}
		\Delta T=\int_{0}^{L}{dx}\left(\frac{1}{c_s}-1 \right)\simeq-\int_{0}^{L}{dx}\,\delta c+O\left (\frac{1}{\Lambda_{UV}}\right )\;,
	\end{eqnarray}
	where $\Lambda_{UV}$ is the cutoff of the EFT, which limits the accuracy of the computation of the phase-shift{, and cannot be larger than the strong-coupling scale, $\Lambda_{UV} \lesssim \Lambda_3$}. It follows that we have a robust prediction of superluminality whenever we find
	\begin{eqnarray}
		\Delta T\lesssim -\frac{1}{\Lambda_{UV} }\;.
	\end{eqnarray}
	This uncertainty is due to lack of precision in the EFT, rather than to the time-localization of the probe. Indeed, one could imagine interferometric measurements of the phase-shift whose precision would not be limited by the probe's frequency.
	
	To keep our analysis as simple as possible, in the following we will consider backgrounds of the form \mbox{$\partial_\mu\phi=\Pi_{\mu\nu}x^\nu$}, with {${\Pi^\mu}_\nu=\partial^{\mu}\partial_\nu {\phi}$} a constant, nonzero, diagonal matrix. As we will discuss in the next section this choice can approximately satisfy the background equations of motion at least on a finite-size region, see Eq.~\eqref{Lmax}, which will be enough to derive a robust estimate of time advance/delay. Moreover, together with the assumption of perturbativity, this choice of background guarantees that propagation will take place on a fixed direction. Note that instead, choosing a background with constant gradient $\partial_\mu\phi=b_\mu$ straightforwardly leads to the positivity bound $c_2>0$ \cite{Adams:2006sv}.
	
	Before closing the section, we remark that one could further study whether the Null Dominant Energy Condition (NDEC) \cite{Curiel2014} is satisfied by the backgrounds considered. NDEC violation would diagnose superluminal transport of energy in the background, further reducing the space of allowed coefficients. However, by inspection of the Stress-Energy tensor, we see that the NDEC violation can only lead to bounds parametrically similar to the ones we derive in this work.

	\section{Superluminality in quartic interactions}
	
	We can start analyzing how the interplay between the different quartic interactions can give raise to superluminality when higher-derivative Galileon-like interactions become larger than lower-derivative shift-symmetric interactions. In practice, let us start by assuming that only $c_2$ and $c_4$ are non zero. We will discuss in a later section how $c_3^2$ contributions enter this picture, obtaining bounds on the same combination of $c_4$ and $c_3^2$ as the one bounded by S-matrix sum rules \cite{Bellazzini:2017fep}. We will work in the perturbative regime in which $(c_4\Pi^2=\epsilon_4\,,\,c_2(\partial\phi)^2 \leq \epsilon_2)\lesssim \epsilon\ll 1 \;${, with $\epsilon_{2,4}$ and $\epsilon$ constants. As we show below, since $(\partial\phi)^2$ is $x$-dependent, the condition of perturbativity on $c_2(\partial\phi)^2$ will only be satisfied in a region of finite size.} At this point, no fixed hierarchy between $c_2$ and $c_4$ is assumed. These conditions, together with limiting the frequency of the perturbation to low enough values, grant that the background will satisfy the Null Energy Condition (NEC) and that no ghost instabilities will appear. With this setup, using the approximate equation of motion for the background $\eta^{\mu\nu}\Pi_{\mu\nu}=O(\epsilon)$, we can solve Eq.~\eqref{eom} and derive the soundspeed deviation from lightspeed. For instance, for a perturbation with momentum $k_\mu=(-c_s k_1,k_1,0,0)$ traveling along $x^\mu=(x_1/c_s,x_1,0,0)+O(\epsilon)$, we find:
	\begin{eqnarray}\label{dc4}
		\delta c= -c_2 x_1^2(\Pi_{00}+\Pi_{11})^2+3c_4(-\Pi_{00}^2+\Pi_{11}^2)+O(\epsilon^2)\,.
	\end{eqnarray}
	Along this trajectory, our perturbative approximation remains valid for a finite length $L_{}$ such that \mbox{$ c_2(\partial\phi)^2 \leq \epsilon_2\ll1\,$}:
	\begin{eqnarray}\label{Lmax}
		L_{}=\dfrac{\epsilon_2^{1/2}}{(c_2|\Pi_{00}^2-\Pi_{11}^2|)^{1/2}}\;.
	\end{eqnarray}
	As mentioned, the term $iD^\mu k_\mu$ {affects the evolution of the amplitude, producing} at leading order in $\epsilon$ an exponential suppression factor $\text{Im}(\omega)x\simeq c_2(\partial\phi)^2\,,$ meaning that restricting our solution to a region of size $L_{}$ also grants that the enhancement or depletion induced by this term can be neglected. Outside of this region we can imagine the background to change in such a way to quench the instability, e.g. turning into $\Pi=0$.
	With this, using Eq.~\eqref{DT}, we find the time-shift accumulated with respect to a lightray:
	\begin{eqnarray} \label{DeltaT-quintic}
		\Delta T &=& \frac{c_2}{3}L_{}^3(\Pi_{00}+\Pi_{11})^2-{3c_4}L_{}(-\Pi_{00}^2+\Pi_{11}^2) \notag \\
		&&+O(L_{}\epsilon^2)\,.
	\end{eqnarray}
	From this expression we can readily derive the positivity bound $c_2>0$, by considering the case in which $\Pi_{00}\sim\Pi_{11}$. To inspect other cases, it is useful to parameterize the background as:
	\begin{eqnarray}
		|-\Pi_{00}^2+\Pi_{11}^2|=\frac{\Lambda^2}{c_2}\quad,&\quad\text{with}\quad\Lambda\sim \frac{\epsilon_2^{1/2}}{L_{}}\;,
	\end{eqnarray}
	and $(\Pi_{00}+\Pi_{11})^2=\alpha\frac{\Lambda^2}{c_2}\,$. Choosing also to identify $\epsilon_4$ with $c_4(-\Pi_{00}^2+\Pi_{11}^2)$, we get:
	\begin{eqnarray}
		\Delta T=L_{}\left (\frac{\alpha}{3}\epsilon_2-3\epsilon_4\right )+O(\epsilon^2)\;.
	\end{eqnarray}
	From this we find that regardless of the sign of $c_4$, there are backgrounds for which one has a robust estimate of superluminality when
	\begin{eqnarray}
		\text{}\;|c_4|{\Lambda^2}\gtrsim c_2\left (\frac{\Lambda}{\Lambda_{UV}}\frac{1}{3\epsilon_2^{1/2}}+\frac{\alpha\epsilon_2}{9}\right )\; ,
	\end{eqnarray}
	which is well compatible with the condition of perturbativity: $\epsilon_2\,,\,\epsilon_4< 1 $. Neglecting the second term and choosing a background such that for instance $\epsilon_2\sim 0.1$, the bound to avoid superluminality becomes:
	\begin{eqnarray}
		\text{}\;| c_4 |\Lambda\Lambda_{UV}\lesssim c_2\quad,\quad \Lambda\sim \frac{1}{3L_{}}\;.
	\end{eqnarray}
	Since $\Lambda$ can be pushed towards $\Lambda_{UV}$ at the expense of perturbative ease, this bound is parametrically equivalent to the one found in \cite{Caron-Huot2021,Bellazzini2022}.

	\section{Superluminality in quintic interactions}
	
	We now consider the main case of interest, in which the quintic interaction has a possibly large coefficient $c_5$. For simplicity, we neglect $c_3$ and $c_4$, keeping only track of the positive $c_2$, and we work in the perturbative regime in which $(c_5\Pi^3=\epsilon_5\,,\,c_2(\partial\phi)^2 \leq \epsilon_2)\lesssim \epsilon\ll 1 \;$, where similarly to the other cases, $\epsilon_5$ is a constant. In this case, evaluating the acoustic metric on the background $\phi$, and again using the approximate equation of motion for the background and considering the same type of perturbation as before we obtain:
	\begin{eqnarray} \label{dc5}
		\delta c &=& -c_2 x_1^2(\Pi_{00}+\Pi_{11})^2+6c_5(\text{tr}[\Pi^2](\Pi_{00}+\Pi_{11}) \notag \\
		&&-2(\Pi_{00}^3+\Pi_{11}^3))+O(\epsilon^2)\;,
	\end{eqnarray}
	where $\text{tr}[\Pi^2]=\Pi_{00}^2+\Pi_{11}^2+\Pi_{22}^2+\Pi_{33}^2$ . From this, we see that as in the previous case the Galileon contribution can lead to superluminal speed regardless of the sign of $c_5$. Similarly to before, $L_{}$ is given by Eq.~\eqref{Lmax} and we have:
	\begin{eqnarray}
		\Delta T=L\left(\frac{\alpha}{3}\epsilon_2-6\epsilon_5\right)+O(\epsilon^2)\;.
	\end{eqnarray}
	Since $\text{tr}[\Pi^2]$ can be as large as $\Lambda_{UV}^2/c_2\,$ without conflicting with perturbativity ($\epsilon_2 \ll 1$), we will find superluminality as soon as:
	\begin{eqnarray}
		|c_5|\Lambda\Lambda_{UV}^{2}\gtrsim c_2^{3/2}\left (\frac{\Lambda}{\Lambda_{UV}}\frac{1}{6\epsilon_2^{1/2}}+\frac{\alpha\epsilon_2}{18}\right )\;.
	\end{eqnarray}
	Similarly to the quartic case, this prediction of superluminality is obtained well in the perturbative regime $\epsilon_5< 1$. Therefore, the requirement of absence of superluminal propagation can be read as the following parametric bound:
	\begin{eqnarray}\label{c5}
		|c_5|\Lambda_{UV}^{3}\lesssim {c_2^{3/2}} \;.
	\end{eqnarray}
	This is a novel two-sided bound on 5-point interactions. Similarly to the case of 4-point interactions, this bound parametrically states that 5-point interactions cannot display weakly broken Galileon symmetry without being in tension with microcausality.

	\section{Field redefinitions and degeneracy of the coefficients}
	
	In the previous sections we have studied cases in which $c_3$ was set to zero. However, we can discuss the cases in which $c_3$ is present by means of a field redefinition. We find that the specific choice:
	\begin{eqnarray}
		\phi &=& \hat{\phi}\,+\,\frac{c_3}{2}\,(\partial\hat{\phi})^2\,+\,\frac{c_3^2}{2}\, \partial\hat{\phi}\!\cdot\!\partial^2\hat{\phi}\!\cdot\!\partial\hat{\phi}\, \notag \\
		&&+ \, \frac{c_3^3}{2}\,\left (\partial\hat{\phi}\!\cdot\!\partial^2\hat{\phi}\!\cdot\!\partial^2\hat{\phi}\!\cdot\!\partial\hat{\phi}+\frac{1}{3}\partial^3\hat{\phi}\!\cdot\!(\partial\hat{\phi})^3\right )\,,
	\end{eqnarray}
	where $\partial^3\hat{\phi}\!\cdot\!(\partial\hat{\phi})^3=\partial^{\mu\nu\rho}\hat{\phi}\,\partial_{\mu}\hat{\phi}\,\partial_{\nu}\hat{\phi}\,\partial_{\rho}\hat{\phi}$ ,
	maps the Galileon-invariant Lagrangian with coefficients $\{c_3,c_4,c_5\}$ to another Galileon-invariant Lagrangian without cubic Galileon term:
	\begin{eqnarray}\label{fredef}
		\left \{\;c_3\;,\;c_4\;,\;c_5\;\right \}\;\to\;\left \{\;0\;,\;c_4-\frac{c_3^2}{2}\;,\;c_5-c_3c_4+\frac{c_3^3}{3}\;\right \}.\,\,\,\,
	\end{eqnarray}
	These combinations are those typically found in 4-pt and 5-pt amplitudes \cite{Bellazzini:2017fep,Kampf2021}. The field redefinition will also give rise to higher derivative 6 and 7 fields interactions, whose contribution is suppressed in the dispersion relation as long as $k$ is small enough. In this new field frame, we will find results similar to those discussed in the previous sections.
	When the quintic interaction is negligible with respect to the quartic we will have, parametrically:
	\begin{eqnarray}
		\left | c_4-\frac{c_3^2}{2}\right |\Lambda_{UV}^2\lesssim c_2\;.
	\end{eqnarray}
	If we exploit the non-renormalization properties of Galileons \cite{Luty:2003vm} to tune the coefficients $c_3$ and $c_4$ so as to make the quartic interaction suppressed and negligible, $c_4=c_3^2/2$, we will have the following parametric bound:
	\begin{eqnarray}
		\left | c_5-\frac{c_3^3}{6}\right |\Lambda_{UV}^3\lesssim c_2^{3/2}\;.
	\end{eqnarray}
	In particular, this bound restricts the coefficients $c_3\,,\,c_4$ even when their contribution to four-point amplitudes is negligible. 
	
	Examining the computation in the original field frame, one finds a contribution linear in $c_3$ to the dispersion relation. Despite this, $c_3$ enters the phase-shift starting at quadratic order. Even more, it can be shown that this observable is sensitive only to the coefficients in the combinations of Eq.~\eqref{fredef}. This originates from the same dynamical redundancy that makes it possible to remove the cubic Galileon interaction with a field redefinition, and confirms the intuition that the phase-shift inherits the field-reparametrization invariance of the S-matrix.
	
	Another way to understand this is by rephrasing the problem of the propagation of the high frequency fluctuations in geometrical terms, provided the term $D^\mu\partial_\mu\varphi$ can be neglected. In this regime, the quadratic action takes the form of a free massless scalar field in curved space with effective metric $G_{\mu\nu}$, such that $\sqrt{-G} \, G^{-1\, \mu\nu} = Z^{\mu\nu}$. The optical path of a scalar perturbation is given by the $G$-null geodesics described by $\Tilde{k}^\mu \equiv G^{-1\, \mu\nu} k_\mu$ :
	\begin{eqnarray}
		\Tilde{k}^\mu \, \nabla^G_\mu \Tilde{k}^\nu = 0,
	\end{eqnarray}
	with $\nabla^G_\mu$ the covariant derivative compatible with $G_{\mu\nu}$ \cite{Heisenberg:2019wjv}. In this description, Shapiro-like time-of-flight differences between neighboring geodesics are associated with a non-vanishing curvature $\mathcal{R}^{\mu}{}_{\nu\rho\sigma}(G)$, which contains no linear in $c_3$ contribution.
	
	One consequence of the degeneracy between the observable effects of $c_3$, $c_4$, and $c_5$ is the existence of a direction in the space of coefficients, $\{c_3,c_3^2/2,c_3^3/6\}$, that is not constrained by superluminality, at least as far as our analysis goes. These ``tuned" Galileons may have coefficients that are parametrically enhanced with respect to the Galileon breaking terms, without tension with microcausality. Despite this, the presence of higher point interactions would presumably break the degeneracy between operators and coefficients in this direction of coupling-space. For instance, as we discuss in the next section, generic Galileon-like higher-point interactions will again lead to time-advances. In addition to this, coupling of the scalar to other fields might also break the degeneracy between the coefficients.
	
	\section{Bounds beyond Galileons}
	
	It is worth considering how the previous analysis applies to other shift-symmetric, but non-Galileon invariant operators. Consider quartic and quintic operators schematically of the form
	\begin{equation}\label{c4c5t}
		\tilde{c}_4 \, (\partial \phi)^2 \Pi^2 \quad,\quad \tilde{c}_5 \, (\partial \phi)^2 \Pi^3\;,
	\end{equation}
	with the same power counting as their Galileon counterparts but with different contractions. While these operators lead to higher-than-second-order equations of motion, in the EFT framework they can be treated consistently at low energies ($E \ll \tilde{c}_4^{-1/6}, \tilde{c}_5^{-1/9}$). For the dispersion relation, this means that any new roots for $\omega(\vec{k})$ lie above the cutoff scale, while the low-energy ones get only perturbatively modified. In what follows we will argue that in this regime, the contributions from $\tilde{c}_{4,5}$ to the soundspeed will be analogous to those of $c_{4,5}$, allowing us to apply unchanged the same bounds from previous sections.

    Let us first consider the four-point interactions by examining all the possible ways we can distribute derivatives in the linearized field equation between the background and the fluctuation. Considering shift-symmetry and the fact that our choice of background is such that $\Pi = const$, we are limited to having either one or two derivatives acting on the background. Moreover, in the perturbative regime in $\tilde{c}_4$, both the background equation of motion and the dispersion relation satisfy, respectively,    
    \begin{eqnarray}
        \Pi^{\mu}{}_{\mu} = O(\epsilon) \quad , \quad k_\mu k^\mu = O(\epsilon) \, .
    \end{eqnarray}
    Therefore, the terms proportional to either of them multiplied times $\tilde{c}_4$ can be safely neglected at leading order. Finally, any terms odd-in-$k_\mu$ are purely imaginary and therefore do not affect the soundspeed itself, but rather are responsible for the suppression of the amplitude of the wave. Therefore, much like the $D^\mu$ term discussed before, their effect can be neglected for sufficiently short distances. Armed with these considerations, it is easy to check that the total contribution to the soundspeed is:
    \begin{eqnarray}
        \tilde{c}_4 \, \Pi^{\mu\alpha} \Pi_{\alpha}{}^\nu \, k_\mu k_\nu + O(\epsilon^2) \, ,
    \end{eqnarray}
    which is the same as in the Galileon case, see Eq.~\eqref{Z}. Therefore, the soundspeed is given just by substituting $c_4 \to \tilde{c}_4$ in Eq.~\eqref{dc4} and we conclude that the bounds discussed before on the quartic Galileon operator also apply to the $\tilde{c}_4$ operators. 
	
	Turning to five-point interactions we can apply similar arguments. In this case, however, apart from Galileon-like contribution there can also be a new relevant $k^4$ contribution to the dispersion relation:
	\begin{eqnarray}\label{tilde5}
		\tilde{c}_5 \left( \Pi^{\mu\nu} \partial^\alpha \phi \, \partial^\beta \phi \right) k_\mu k_\nu k_\alpha k_\beta+O(\epsilon^2)\;.
	\end{eqnarray}
	Considering a background as in the previous sections, this contribution affects the soundspeed in a way similar to the quintic Galileon, Eq.~\eqref{dc5}, up to a factor $(k L)^2 > 1$. This implies a bound on $\tilde{c}_5$ at least as good as the one on $c_{5}$, if not stronger. Since no physical scale separates $k$ from $\Lambda$, we expect the bound to remain parametrically similar to Eq.~\eqref{c5}. From the S-matrix perspective, the fact that the bounds on $\tilde{c}_{4,5}$ are the same as the ones for $c_{4,5}$ follows from the fact that there is only one contribution to the on-shell 4-point and 5-point scattering amplitudes at this order in derivatives.
	
	Before concluding, we consider the case of Galileon-like operators with higher number of fields, such as:
	\begin{eqnarray}
		c_n(\partial\phi)^2\Pi^{n-2}\;,
	\end{eqnarray} 
	with $n \geq 6$. In the presence of $c_2(\partial\phi)^4$, $c_n$ will give a contribution to the time-shift of order 
	\begin{eqnarray}
		c_nL_{}\frac{\Lambda^{n-2}}{c_2^{(n-2)/2}}\left[ 1+(kL_{})^2 \right]+O(\epsilon^2)\;.
	\end{eqnarray} 
	Then, if the combination of constant background $\Pi^{n-2}$ can consistently get both signs, we will get the following two-sided bounds:
	\begin{eqnarray}\label{cn}
		|c_n|\Lambda^{n-3}\Lambda_{UV}\lesssim c_2^{(n-2)/2}\;.
	\end{eqnarray}

	\section{Discussion}
	
	In this letter we studied how the presence of superluminal propagation at low energies is reflected in the coefficients of 5-point interactions. In the case of a self-interacting scalar field, we derived two-sided bounds on these coefficients by computing the time advance accumulated by a scalar perturbation traveling through a region with a given background. Our analysis is made particularly straightforward by considering propagation on a finite region over which both the background and the dynamics of perturbations have a simple description. This simplicity comes at the expense of the sharpness of our bounds, which despite being robust are only parametrically precise. For concreteness, we derived explicitly the bound on the quintic Galileon interaction, Eq.~\eqref{c5}, extending them later to other shift-symmetric 5-point scalar self-interactions having the same number of derivatives, Eq.~\eqref{c4c5t}. These bounds state superluminality will appear when these quintic interactions are large compared to the shift symmetric quartic ones. If we write $c_2\sim1/\Lambda_2^4\,,\,c_5\sim1/\Lambda_3^9$, then the bound can be read parametrically as stating that $\Lambda_3^9\gtrsim\Lambda_2^6\Lambda_{UV}^3\,$. This means that a situation of (weakly broken) Galileon invariance, in which $\Lambda_3$ is low compared to the other scales, is in tension with microcausality at the level of 5-point interactions.
	
	Besides 5-point interactions, we considered 4-point interactions and showcased how our simplified analysis can reproduce the existing constraints arising from S-matrix sum rules in this context. Moreover, we argued that Galileon-like interactions with higher number of fields should be compatible with microcausality only when Eq.~\eqref{cn} holds. 
	
	Although our analysis does not cover all higher-derivative operators that respect Galileon symmetry, it still leads to a conclusive statement about Galileon symmetry and high-derivative interactions at low energy. Indeed, our five-point bound removes the possibility of having dominant ghost-free Galileon interactions, unless nature admits faster-than-light propagation. This means that there is no way to produce significant nonlinearities within the EFT regime of validity, provided a stable Minkowski vacuum is allowed.
	
	In conclusion, it will be interesting to employ the simple analytic strategy presented here to investigate other classes of interactions, both at higher-points and in contexts more relevant for phenomenology.

    \vspace{-0.25cm}
	
	\begin{acknowledgments}
		\section{Acknowledgments}
		The authors would like to thank B. Bellazzini, D. E. \mbox{Kaplan}, D. Kosmopoulos, S. Rajendran, S. Ramazanov, J. Serra, G. Trenkler, A. Vikman for useful discussions, and especially I. \mbox{Sawicki} and E. Trincherini for helpful discussions and comments on the manuscript.
		The work of F.S. was supported by the U.S. Department of Energy (DOE), Office of Science, National Quantum Information Science Research Centers, Superconducting Quantum Materials and Systems Center (SQMS) under Contract No. DE-AC02-07CH11359 as well as by the Simons Investigator Award No. 827042 (P.I.: Surjeet Rajendran). The work of L.G.T. was supported by European Union (Grant No. 101063210). We also thank the hospitality of the Centro de Ciencias de Benasque Pedro Pascual during the initial stages of this work.
	\end{acknowledgments}

    \vspace{-0.25cm}

    \bibliographystyle{unsrt}
	\bibliography{refs}

\begin{thebibliography}{10}

\bibitem{Pham:1985}
T.~N. Pham and Tran~N. Truong.
\newblock {Evaluation of the Derivative Quartic Terms of the Meson Chiral
  Lagrangian From Forward Dispersion Relation}.
\newblock {\em Phys. Rev. D}, 31:3027, 1985.

\bibitem{Adams:2006sv}
Allan Adams, Nima Arkani-Hamed, Sergei Dubovsky, Alberto Nicolis, and Riccardo
  Rattazzi.
\newblock {Causality, analyticity and an IR obstruction to UV completion}.
\newblock {\em JHEP}, 10:014, 2006.

\bibitem{Arkani-Hamed2021a}
Nima Arkani-Hamed, Tzu-Chen Huang, and Yu-Tin Huang.
\newblock {The EFT-Hedron}.
\newblock {\em JHEP}, 05:259, 2021.

\bibitem{Bros1964}
J.~Bros, H.~Epstein, and Vladimir~Jurko Glaser.
\newblock {Some rigorous analyticity properties of the four-point function in
  momentum space}.
\newblock {\em Nuovo Cim.}, 31:1265--1302, 1964.

\bibitem{Mizera2024}
Sebastian Mizera.
\newblock {Physics of the analytic S-matrix}.
\newblock {\em Phys. Rept.}, 1047:1--92, 2024.

\bibitem{Nicolis:2009qm}
Alberto Nicolis, Riccardo Rattazzi, and Enrico Trincherini.
\newblock {Energy's and amplitudes' positivity}.
\newblock {\em JHEP}, 05:095, 2010.
\newblock [Erratum: JHEP 11, 128 (2011)].

\bibitem{Bellazzini2017}
Brando Bellazzini.
\newblock {Softness and amplitudes\textquoteright{} positivity for spinning
  particles}.
\newblock {\em JHEP}, 02:034, 2017.

\bibitem{Bellazzini:2017fep}
Brando Bellazzini, Francesco Riva, Javi Serra, and Francesco Sgarlata.
\newblock {Beyond Positivity Bounds and the Fate of Massive Gravity}.
\newblock {\em Phys. Rev. Lett.}, 120(16):161101, 2018.

\bibitem{Bellazzini:2020cot}
Brando Bellazzini, Joan Elias~Mir\'o, Riccardo Rattazzi, Marc Riembau, and
  Francesco Riva.
\newblock {Positive moments for scattering amplitudes}.
\newblock {\em Phys. Rev. D}, 104(3):036006, 2021.

\bibitem{Tolley2021}
Andrew~J. Tolley, Zi-Yue Wang, and Shuang-Yong Zhou.
\newblock {New positivity bounds from full crossing symmetry}.
\newblock {\em JHEP}, 05:255, 2021.

\bibitem{Caron-Huot2021}
Simon Caron-Huot and Vincent Van~Duong.
\newblock {Extremal Effective Field Theories}.
\newblock {\em JHEP}, 05:280, 2021.

\bibitem{Bellazzini2022}
Brando Bellazzini, Marc Riembau, and Francesco Riva.
\newblock {IR side of positivity bounds}.
\newblock {\em Phys. Rev. D}, 106(10):105008, 2022.

\bibitem{Caron-Huot:2021rmr}
Simon Caron-Huot, Dalimil Mazac, Leonardo Rastelli, and David Simmons-Duffin.
\newblock {Sharp Boundaries for the Swampland}.
\newblock {\em JHEP}, 07:110, 2021.

\bibitem{Chiang:2021ziz}
Li-Yuan Chiang, Yu-tin Huang, Wei Li, Laurentiu Rodina, and He-Chen Weng.
\newblock {Into the EFThedron and UV constraints from IR consistency}.
\newblock {\em JHEP}, 03:063, 2022.

\bibitem{Caron-Huot2022}
Simon Caron-Huot, Yue-Zhou Li, Julio Parra-Martinez, and David Simmons-Duffin.
\newblock {Causality constraints on corrections to Einstein gravity}.
\newblock 1 2022.

\bibitem{Hong:2023zgm}
Dong-Yu Hong, Zhuo-Hui Wang, and Shuang-Yong Zhou.
\newblock {Causality bounds on scalar-tensor EFTs}.
\newblock {\em JHEP}, 10:135, 2023.

\bibitem{Bros1972}
J.~Bros, V.~Glaser, and H.~Epstein.
\newblock {Local analyticity properties of the n particle scattering
  amplitude}.
\newblock {\em Helv. Phys. Acta}, 45:149--181, 1972.

\bibitem{CaronHuot2023a}
Simon Caron-Huot, Mathieu Giroux, Holmfridur~S. Hannesdottir, and Sebastian
  Mizera.
\newblock {What can be measured asymptotically?}
\newblock 8 2023.

\bibitem{CaronHuot2023}
Simon Caron-Huot, Mathieu Giroux, Holmfridur~S. Hannesdottir, and Sebastian
  Mizera.
\newblock {Crossing beyond scattering amplitudes}.
\newblock 10 2023.

\bibitem{Bros1986}
J.~Bros.
\newblock {Derivation of asymptotic crossing domains for multiparticle
  processes in axiomatic quantum field theory: A general approach and a
  complete proof for 2 ---\ensuremath{>} 3 particle processes}.
\newblock {\em Phys. Rept.}, 134:325, 1986.

\bibitem{Chandrasekaran2018}
Venkatesa Chandrasekaran, Grant~N. Remmen, and Arvin Shahbazi-Moghaddam.
\newblock {Higher-Point Positivity}.
\newblock {\em JHEP}, 11:015, 2018.

\bibitem{Camanho:2014apa}
Xian~O. Camanho, Jose~D. Edelstein, Juan Maldacena, and Alexander Zhiboedov.
\newblock {Causality Constraints on Corrections to the Graviton Three-Point
  Coupling}.
\newblock {\em JHEP}, 02:020, 2016.

\bibitem{CarrilloGonzalez2022}
Mariana Carrillo~Gonzalez, Claudia de~Rham, Victor Pozsgay, and Andrew~J.
  Tolley.
\newblock {Causal effective field theories}.
\newblock {\em Phys. Rev. D}, 106(10):105018, 2022.

\bibitem{Serra2022}
Francesco Serra, Javi Serra, Enrico Trincherini, and Leonardo~G. Trombetta.
\newblock {Causality constraints on black holes beyond GR}.
\newblock {\em JHEP}, 08:157, 2022.

\bibitem{CarrilloGonzalez2023}
Mariana Carrillo~Gonz\'alez, Claudia de~Rham, Sumer Jaitly, Victor Pozsgay, and
  Anna Tokareva.
\newblock {Positivity-causality competition: a road to ultimate EFT consistency
  constraints}.
\newblock 7 2023.

\bibitem{Greene:2022uyf}
Brian Greene, Daniel Kabat, Janna Levin, and Massimo Porrati.
\newblock {Back to the future: Causality on a moving braneworld}.
\newblock {\em Phys. Rev. D}, 107(2):025016, 2023.

\bibitem{Nicolis2009b}
Alberto Nicolis, Riccardo Rattazzi, and Enrico Trincherini.
\newblock {The Galileon as a local modification of gravity}.
\newblock {\em Phys. Rev. D}, 79:064036, 2009.

\bibitem{Maiani:1997pd}
L.~Maiani and M.~Testa.
\newblock {Unstable systems in relativistic quantum field theory}.
\newblock {\em Annals Phys.}, 263:353--367, 1998.

\bibitem{Curiel2014}
Erik Curiel.
\newblock {A Primer on Energy Conditions}.
\newblock {\em Einstein Stud.}, 13:43--104, 2017.

\bibitem{Kampf2021}
Karol Kampf, Jiri Novotny, Filip Preucil, and Jaroslav Trnka.
\newblock {Multi-spin soft bootstrap and scalar-vector Galileon}.
\newblock {\em JHEP}, 07:153, 2021.

\bibitem{Luty:2003vm}
Markus~A. Luty, Massimo Porrati, and Riccardo Rattazzi.
\newblock {Strong interactions and stability in the DGP model}.
\newblock {\em JHEP}, 09:029, 2003.

\bibitem{Heisenberg:2019wjv}
Lavinia Heisenberg and Christian~F. Steinwachs.
\newblock {Geometrized quantum Galileons}.
\newblock {\em JCAP}, 02:031, 2020.

\end{thebibliography}
	
\end{document}